\documentclass[nofootinbib,showkeys,showpacs,tightenlines,fleqn]{revtex4}  
\usepackage{graphicx}
\usepackage{dcolumn}
\newcommand{\be}{\begin{equation}}
\newcommand{\ee}{\end{equation}}
\newcommand{\bee}{\begin{eqnarray}}
\newcommand{\eee}{\end{eqnarray}}
\newcommand{\eq}{\end{quote}}
\newcommand{\nn}{\nonumber}

\begin{document}
\title{QCD condensates with flavor SU(3) symmetry breaking}  
\author{Seung-il Nam}
\email{sinam@pusan.ac.kr}
\affiliation{Department of
Physics and Nuclear Physics \& Radiation Technology Institute (NuRI),  
Pusan National University, Busan 609-735, Republic of Korea}
\author{Hyun-Chul Kim}
\email{hchkim@pusan.ac.kr}
\affiliation{Department of
Physics and Nuclear Physics \& Radiation Technology Institute (NuRI),
Pusan National University, Busan 609-735, Republic of Korea}
\date{May 2006}
\begin{abstract}
We investigate the effects of flavor SU(3)-symmetry breaking on the QCD condensates, based on the nonlocal 
effective chiral action from the instanton vacuum, focusing on the
quark-gluon mixed one.  We employ two
different parameterizations for the dependence of the
momentum-dependent dynamical quark mass on the strange current quark
mass. We obtain the ratios of the mixed condensates: 
$[\langle\bar{s}\sigma_{\mu\nu}G^{\mu\nu}s
\rangle/\langle\bar{u}\sigma_{\mu\nu}G^{\mu\nu}u\rangle]^{1/5}=0.87$.  It
turns 
out that the dimensional parameter 
$m^2_0=\langle\bar{q} \sigma_{\mu\nu}G^{\mu\nu}q\rangle/
\langle\bar{q}q\rangle =1.60\sim 1.92\,{\rm GeV}^2$. 
\end{abstract}
\maketitle
\section{Introduction}
In this proceeding~\footnote{This proceeding is prepared for the international
workshop, ``Hadrons at finite density'' (HFD06) at YITP in Japan 20$\sim$22
Feb 2006.}, we discuss the effect of flavor
SU(3)-symmetry breaking focusing on the quark-gluon mixed condensate, based on
the instanton liquid model for the QCD 
vacuum~\cite{Diakonov:1985eg,Diakonov:2002fq}. The model was later extended
by  
introducing the current quark masses~\cite{Musakhanov:1998wp,Musakhanov:vu}.
Since we are interested in the 
effect of explicit flavor SU(3)-symmetry breaking, we follow the
formalism in Ref.~\cite{Musakhanov:1998wp}.  Though the mixed
quark-gluon condensate was already studied in the instanton vacuum 
~\cite{Polyakov:1996kh}, explicit SU(3)-symmetry breaking was not
considered.  Hence, we extend the work of
Ref.~\cite{Polyakov:1996kh}, focussing on the effect of flavor   
SU(3)-symmetry breaking on the QCD vacuum.  Furthermore, we take into
account two different parameterizations for the dependence of the
dynamical quark mass on the current quark mass $m_f$ so that we can
examine the effects of the current quark mass very in detail.  We observed
that, with a proper choice of the $m_f$ dependence of the 
dynamical quark mass~\cite{Pobylitsa:1989uq}, the gluon condensate is
independent of $m_f$~\cite{Nam:2006ng}.  The corresponding results are
summarized as 
follows: The ratio $[{\langle
\bar{s}\sigma_{\mu\nu}G^{\mu\nu}s\rangle}/{\langle
\bar{u}\sigma_{\mu\nu}G^{\mu\nu}u\rangle}]^{1/5}  =
0.87$, $m_{0,{\rm u}}^2=1.60\,{\rm GeV}^2$, and $m_{0,{\rm
    s}}^2=1.84\,{\rm GeV}^2$, with isospin symmetry assumed. This proceeding
is based on our previous work of Ref.~\cite{Nam:2006ng} in which the gluon and
quark condensates are were also discussed in addition to the mixed one.     

\section{Formalism}
Now, we consider the mixed condensate, 
$\langle\bar{q}\sigma_{\mu\nu}G^{\mu\nu}q\rangle$ in our framework.  Actually,
the local 
operator inside this condensate corresponds to the quark-gluon
interaction of a Yukawa type.  However, in the present work, the gluon field
strength ($G_{\mu\nu}$) can be expressed in terms of the quark-instanton
interaction~\cite{Polyakov:1996kh}.  First, the one
flavor quark and one instanton interaction can be
rewritten as a function of space-time coordinates $x$ and color
orientation matrix $U$:   
\bee
Y_{\pm,1}(x,U)=\int\frac{d^4k}{(2\pi)^4}\frac{d^4p}{(2\pi)^4}[
2\pi\rho F(k\rho)][2\pi\rho F(p\rho)]e^{-ix\cdot(k-p)}\left[U^{\alpha}_{i'}(U^{j'}_{\beta})^{\dagger}\epsilon^{ii'}
\epsilon_{jj'}\right]\left[i\psi^{\dagger}(k)_{\alpha i}
\frac{1\pm\gamma_5}{2}\psi(p)^{\beta j}\right],
\label{vertex2}
\eee 
where the form factor, $F(k\rho)$, is defined as follows:
\bee
F (k\rho)
=2t\left[I_0(t)K_1(t)-I_1(t)K_0(t)-\frac{1}{t}I_1(t)K_1(t)\right]. 
\label{FF1}
\eee
$\bar{\rho}$ is set to be $1/600$ MeV$^{-1}$. We assume the
$\delta$-function-type instanton distribution. We  
define then the field strength $G^a_{\mu\nu}$ in terms of the instanton
configuration as a function of $I$($\bar{I}$) position and orientation
matrix $U$: 
\bee
G^a_{\pm\mu\nu}(x,x',U)=\frac{1}{2}\left[\lambda^aU\lambda^b
U^{\dagger}\right]G^b_{\pm\mu\nu}(x'-x). 
\label{field}
\eee 
$G^b_{\pm\mu\nu}(x'-x)$ stands for the field strength consisting of a
certain instanton configuration.  Using Eqs.~(\ref{vertex2}) and
(\ref{field}), we define the field strength in terms of the 
quark-instanton interaction: 
\bee
\hat{G}^a_{\pm\mu\nu}=\frac{iN_cM}{4\pi\bar{\rho}^2}\int d^4x
\int dU G^a_{\pm\mu\nu}(x,x',U)Y_{\pm,1}(x,U) 
\eee  
Following the method in Ref.~\cite{Polyakov:1996kh},
we finally obtain the quark-gluon mixed condensate as follows: 
\bee
\langle\bar{q}\sigma_{\mu\nu}G^{\mu\nu}q\rangle=2N_c\bar{\rho}^2\int
\frac{d^4k_1}{(2\pi)^4}\int\frac{d^4k_2}{(2\pi)^4}\frac{
\sqrt{M(k_1)M(k_2)}G(k_1,k_2)N(k_1,k_2)}{[k^2_1+[m_f+M(k_1)]^2][
k^2_2+[m_f+M(k_2)]^2]}, 
\label{mc}
\eee
where $G(k_1,k_2)$ and $N(k_1,k_2)$ are defined as follows:
\bee
G(k_1,k_2)&=&32\pi^2\left[\frac{K_0(t)}{2}+\left\{\frac{4K_0(t)}{t^2}
+\left(\frac{2}{t}+\frac{8}{t^3}\right)K_1(t)-\frac{8}{t^4}\right\}\right],
\nn\\ 
N(k_1,k_2)&=&\frac{1}{(k_1-k_2)^2}\left[8k^2_1k^2_2-6(k^2_1+k^2_2)k_1\cdot
  k_2+4(k_1\cdot k_2)^2\right]
\eee
with $t=|k_1-k_2|\bar{\rho}$.  If we
consider for arbitrary $N_f$ the mixed condensate, the situation may be
somewhat different from the cases of the gluon and quark condensates.
We note that, though the mixed condensate has been calculated for the
case of $N_f=1$, the same formula of Eq.~(\ref{mc}) still holds for
each flavor with arbitrary $N_f$ as discussed
previously~\cite{Polyakov:1996kh}.   

As indicated in Refs.~\cite{Musakhanov:1998wp,Kim:2005jc}, the
dynamical quark mass is also a function of the current quark
mass.  Thus, we redefine the dynamical quark mass as a 
function of momentum and current quark mass, $m_f$:
\bee
M(k)\to M(k,m_f)=M_0f(m_f)F^2(k\bar{\rho}),
\eee 
where we set $M_0$ to be $350$ MeV. In order to take into account the
dependence of the dynamical quark 
mass on $m_f$, we introduce the $m_f$-dependent correction factor
$f(m_f)$~\cite{Musakhanov:1998wp,Kim:2005jc}.  We consider here two  
different parameterizations for $f (m_f)$ ($A$ and $B$) of $f(m_f)$ to
see theoretical ambiguities: 
\bee
f_A(m_f)=\sqrt{1+\frac{m^2_f}{c^2_A}}-\frac{m_f}{c_A},\;\;\;
f_B(m_f)=1+\frac{c_B}{M_0},\;\;\;  f_C (m_f) = 1.
\label{mfc}
\eee
The correction factor $f_A(m_f)$ is derived by
Pobylitsa~\cite{Pobylitsa:1989uq} by expanding the quark propagator in
the instanton background in terms of the large $N_c$ limit.  The
parameter $c_A$ is set to $0.198$ GeV.  The $m_f$ correction factor 
$f_B(m_f)$, was introduced in Ref.~\cite{Musakhanov:1998wp} by using  
the saddle-point equation and its expansion in the current quark mass
$m_f$.  We use the value of $c_B=-0.5\,m_f$ as proposed in
Ref.~\cite{Musakhanov:1998wp}.  $f_C$ is for the $M$ without $m_f$
dependence.   
\section{Numerical results}
\begin{figure}[t]
\begin{center}
\begin{tabular}{cc}
\includegraphics[width=8cm]{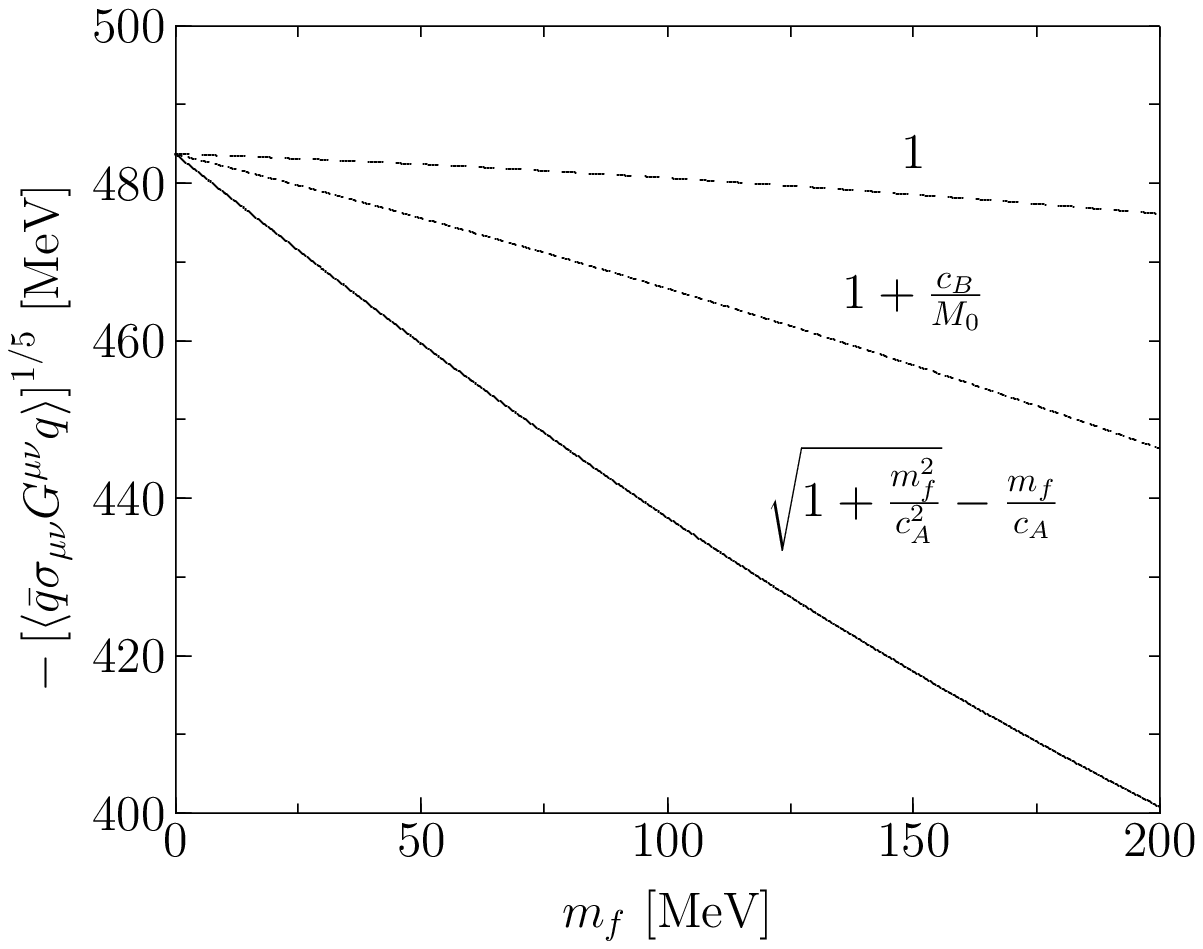}
\includegraphics[width=8cm]{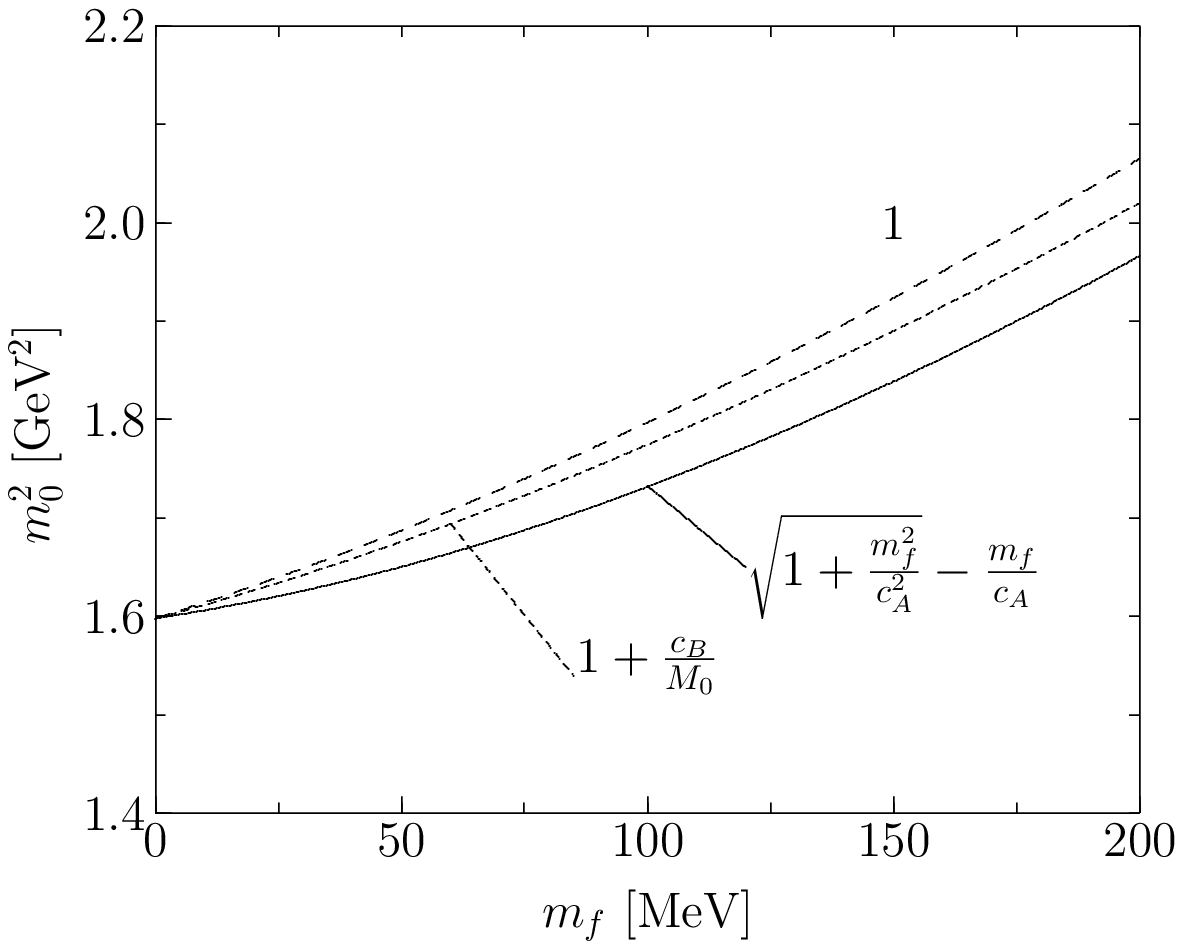}
\end{tabular}
\caption{Left: Quark-gluon mixed  condensate as a function of the current
quark mass $m_f$.  Right: $m^2_0$ as a function of the current quark mass.
The solid curve draws the mixed condensate for 
$f_A(m_f)$ in Eq.~(\ref{mfc}), the dashed one corresponds to that
without any $f(m_f)$, and the dotted one plots that for $f_B(m_f)$.
We used $c_A=0.198$ GeV and $c_B=-0.5\,m_f$ GeV.}
\label{fig1} 
\end{center}
\end{figure}

In the right panel of Fig.~\ref{fig1}, we draw the results of the quark-gluon
mixed 
condensate as functions of the $m_f$. The curves of the mixed condensate
decrease as $m_f$ increases.    
We also list the values of the mixed condensate for the up and 
strange quarks in Table.~\ref{table1}. 
\begin{table}[ht]
\begin{tabular}{c|ccc}
&$A$&$B$&$C$\\
\hline
$\langle\bar{u}\sigma_{\mu\nu}G^{\mu\nu}u\rangle$&$-$481$^5$&$-$483$^5$&$-$484$^5$\\
$\langle\bar{s}\sigma_{\mu\nu}G^{\mu\nu}s\rangle$&$-$418$^5$&$-$475$^5$&$-$483$^5$
\end{tabular}
\caption{Quark-gluon mixed condensates for $m_u=5$ and $m_s=150$ MeV
 [MeV$^5$].} 
\label{table1}
\end{table}

Now, we consider the effect of flavor SU(3)-symmetry breaking by
calculating the ratios between the nonstrange condensates and the
strange ones.  As already discussed, the ratio of the gluon
condensates remains unity for the $m_f$ correction factor $f_A$. The
calculated ratios are given by;
$[\langle\bar{s}\sigma_{\mu\nu}G^{\mu\nu}s\rangle/\langle\bar{u} 
\sigma_{\mu\nu}G^{\mu\nu}u\rangle]^{1/5}=0.87\sim1.00$.  Being compared to 
Refs.~\cite{Beneke:1992ba,Aladashvili:1995zj,Khatsimovsky:1987bb,Ovchinnikov:1988gk,Braun:2004vf},
our results are in good agreement each other.

We take into account a dimensional quantity $m_0^2$ defined as the ratio
between the mixed and quark condensates: 
\bee
m^2_0=\langle\bar{q}\sigma_{\mu\nu}G^{\mu\nu}q\rangle/\langle\bar{q}q\rangle,
\eee  
which is an important input for general QCD sum rule calculations. We
draw $m_0^2$ in the left panel of Fig.~\ref{fig1} as a function of $m_f$ and list the values of
$m^2_0$ in Table.~\ref{table4} in which the quark condensates are computed in
the same framework.  

\begin{table}[ht]
\begin{tabular}{c|ccc}
&$A$&$B$&$C$\\
\hline
$[\langle\bar{s}\sigma_{\mu\nu}G^{\mu\nu}s\rangle/\langle\bar{u}
\sigma_{\mu\nu}G^{\mu\nu}u\rangle]^{1/5} $&0.87&0.98&1.00\\ 
$m^2_{0,u}=\langle\bar{u}\sigma_{\mu\nu}G^{\mu\nu}u\rangle/
\langle\bar{u}u\rangle$&1.60 GeV$^2$&1.60 GeV$^2$&1.60 GeV$^2$\\
$m^2_{0,s}=\langle\bar{s}\sigma_{\mu\nu}G^{\mu\nu}s\rangle/
\langle\bar{s}s\rangle$&1.84 GeV$^2$&1.89 GeV$^2$&1.92 GeV$^2$\\
\end{tabular}
\caption{The ratios of the condensates for the different types of the
  $m_f$ correction factors. $m_u=5$ MeV and $m_s=150$ MeV are used.}
\label{table4}
\end{table} 
The value of $m_0^2$ increases as $m_f$ does, which implies that the
mixed condensate is less sensitive to the $m_f$ than the quark
condensate.  The values of $m_0^2$ are in the range of $1.84\, {\rm
  GeV}^2$ for the strange quark and of $1.60 \,{\rm GeV}^2$ for the up
quark.  Thus, the strange $m_{0,s}^2$ turns out to be larger than the
nonstrange $m_{0,u}^2$ by about $15\,\%$.      
\section{Summary and Conclusion}
In the present work, we investigated the various QCD vacuum
condensates within the framework of the instanton liquid model,
emphasizing the effects of flavor SU(3)-symmetry breaking.  The
modified improved action elaborated by Musakhanov was used for
this purpose.  In the modified improved action, the current 
quark mass appeared explicitly in the denominator of the quark
propagator as well as in the dynamical quark mass.  Thus, we were able
to take into account the current quark mass effects to the QCD 
condensates.  In order to consider the $m_s$ dependence of the
dynamical quark mass, we employed two different 
types of the correction factors, $f_A$ and $f_B$ ($f_C$ for the case
without the correction). $f_A$ arises from the resummation of the QCD
planar loops in the large $N_c$ limit~\cite{Pobylitsa:1989uq}, while
$f_B$ is a simple parameterization of the current quark mass
correction suggested by Musakhanov~\cite{Musakhanov:1998wp}.

The mixed condensates were calculated with these
correction factors.  The results were consistent with those from
other model calculations as well as phenomenological values. 
In particular,  The ratios of the condensates between the strange and
up quarks were also investigated:
$[\langle\bar{s}\sigma_{\mu\nu}G^{\mu\nu}s\rangle/\langle\bar{u}
\sigma_{\mu\nu}G^{\mu\nu}u\rangle]^{1/5}$ and
$\langle\bar{q}\sigma_{\mu\nu}G^{\mu\nu}q\rangle/ \langle
\bar{q}q\rangle$.  It turned 
out that the results are again compatible with other theoretical
calculations.  The dimensional quantity $m^2_{0}$ was also studied:
$m^2_{0,u}=1.6\,{\rm GeV}^2$ and $m^2_{0,s}=1.84\,{\rm GeV}^2$.  In
general, the quark and mixed condensates decrease as the current  
quark mass increases.  However, the $m_0^2$ increases as the current
quark mass does, which indicates that the mixed condensate is less
sensitive to the current quark mass than the quark condensate. More details
can be found in Ref.~\cite{Nam:2006ng}. 

\section*{Acknowledgments}
The present work is supported by the Korean Research Foundation
(KRF--2003--070--C00015).  The authors are grateful to
M.~M.~Musakhanov and Y.~Kwon for fruitful discussions.   
S.~I.~N.~would like to thank A.~Hosaka and RCNP for the support during the
YITP workshop ``Hadrons at finite density'' at Kyoto, Japan 20$\sim$22 Feb
2006.    
\bibliographystyle{ws-procs9x6}
\bibliography{ws-pro-sample}

\begin{thebibliography}{99}
\bibitem{Diakonov:1985eg}
D.~Diakonov and V.~Y.~Petrov,Nucl.\ Phys.\ B {\bf 272}, 457 (1986).
\bibitem{Diakonov:2002fq}
D.~Diakonov, Prog.\ Part.\ Nucl.\ Phys.\  {\bf 51}, 173 (2003).
\bibitem{Musakhanov:1998wp}
M.~Musakhanov, Eur.\ Phys.\ J.\ C {\bf 9}, 235 (1999).
\bibitem{Musakhanov:vu}
M.~Musakhanov, Nucl.\ Phys.\ A {\bf 699}, 340 (2002).
\bibitem{Polyakov:1996kh}
M~.~V.~Polyakov and C.~Weiss, Phys.\ Lett.\ B {\bf 387}, 841 (1996). 
\bibitem{Pobylitsa:1989uq}
P.~V.~Pobylitsa, Phys.\ Lett.\ B {\bf 226}, 387 (1989).
\bibitem{Nam:2006ng}
S.~i.~Nam and H.~Ch-.~Kim, arXiv:hep-ph/0605041.
\bibitem{Kim:2005jc}
H.-Ch.~Kim, M.~M.~Musakhanov and M.~Siddikov, Phys.\ Lett.\ B {\bf 633}, 201
(2006).  
\bibitem{Schafer:1996wv}
T.~Schafer and E.~V.~Shuryak, Rev.\ Mod.\ Phys.\  {\bf 70}, 323 (1998).
\bibitem{Jamin:2002ev}
M.~Jamin, Phys.\ Lett.\ B {\bf 538}, 71 (2002).
\bibitem{Dosch:1988vv}
H.~G.~Dosch, M.~Jamin and S.~Narison, Phys.\ Lett.\ B {\bf 220}, 251 (1989).
\bibitem{DiGiacomo:2004ff}
A.~Di Giacomo and Y.~A.~Simonov, Phys.\ Lett.\ B {\bf 595}, 368
(2004). 
\bibitem{Beneke:1992ba}
M.~Beneke and H.~G.~Dosch, Phys.\ Lett.\ B {\bf 284}, 116 (1992).
\bibitem{Aladashvili:1995zj}
K.~Aladashvili and M.~Margvelashvili, Phys.\ Lett.\ B {\bf 372}, 299 (1996). 
\bibitem{Khatsimovsky:1987bb}
V.~M.~Khatsimovsky, I.~B.~Khriplovich and A.~R.~Zhitnitsky, Z.\ Phys.\ C {\bf
  36}, 455 (1987). 
\bibitem{Ovchinnikov:1988gk}
A.~A.~Ovchinnikov and A.~A.~Pivovarov, Sov.\ J.\ Nucl.\ Phys.\  {\bf 48}, 721
(1988) [Yad.\ Fiz.\  {\bf 48}, 1135 (1988)].
\bibitem{Braun:2004vf}
V.~M.~Braun and A.~Lenz, Phys.\ Rev.\ D {\bf 70}, 074020 (2004).

\end{thebibliography}

\end{document}